# Resource Allocation for Secure Communications in Cooperative Cognitive Wireless Powered Communication Networks

Ding Xu, *Member, IEEE,* and Qun Li, *Member, IEEE*

*Abstract*—We consider a cognitive wireless powered communication network (CWPCN) sharing the spectrum with a primary network who faces security threats from eavesdroppers (EAVs). We propose a new cooperative protocol for the wireless powered secondary users (SU) to cooperate with the primary user (PU). In the protocol, the SUs first harvest energy from the power signals transmitted by the cognitive hybrid access point during the wireless power transfer (WPT) phase, and then use the harvested energy to interfere with the EAVs and gain transmission opportunities at the same time during the wireless information transfer (WIT) phase. Taking the maximization of the SU ergodic rate as the design objective, resource allocation algorithms based on the dual optimization method and the block coordinate descent method are proposed for the cases of perfect channel state information (CSI) and collusive/non-collusive EAVs under the PU secrecy constraint. More PU favorable greedy algorithms aimed at minimizing the PU secrecy outage probability are also proposed. We furthermore consider the unknown EAVs' CSI case and propose an efficient algorithm to improve the PU security performance. Extensive simulations show that our proposed protocol and corresponding resource allocation algorithms can not only let the SU gain transmission opportunities but also improve the PU security performance even with unknown EAVs' CSI.

## I. Introduction

COGNITIVE radio (CR) has been widely studied in recent years due to its capability of improving spectrum utilization [1]. In CR networks, there are two kinds of users, one is called secondary user (SU) who does not have its own spectrum, and the other one is called primary user (PU) who owns licensed spectrum. Basically, there are three main CR paradigms for the SU to utilize the spectrum licensed to the PU: interweave, underlay and overlay [2]. In interweave CR networks, the SU must sense the spectrum status and utilize the spectrum when the spectrum is sensed as vacant. In underlay CR networks, the SU can utilize the spectrum as long as the interference caused to the PU is acceptable for the PU. These two CR paradigms do not provide incentives for the PU to allow the SU to utilize its spectrum. In contrast to interweave and underlay CR networks, in overlay CR networks, the transmission of the PU is facilitated from the cooperation from the SU. By doing so, the PU has motivation for allowing the SU to utilize the licensed spectrum.

Security is an important requirement for wireless networks, and there is no exception for the PU networks. In particular, we are interested in one kind of security threats to the PU networks, i.e., eavesdropping from malicious users who try to decode messages broadcasted from the PU. In this context, secrecy rate [3], [4], defined as the achievable non-negative rate of the perfectly secure information sent from the source to the destination with the presence of the eavesdroppers (EAVs), is usually adopted to measure the physical-layer security performance. However, only zero or trivial secrecy rate can be achieved by the PU when the source-to-EAV channel is strong. Generally, the SU can act as a cooperative relay or a friendly jammer to improve the PU security performance [5].

Meanwhile, wireless power transfer (WPT) has greet potential to become a perpetual energy source for wireless equipments by harvesting ambient radio signals [6], [7]. For example, a power station transmitting tens of watts can support maximum WPT distance in the range of $3 - 15$ meter for powering smart phones and tablet computers [8]. Thus, wireless powered communication networks (WPCNs), in which wireless equipments are powered by WPT, have received a lot of attention . A typical transmission protocol for the WPCNs is the so-called "harvest-then-transmit" protocol [9], where in each transmission time block, users first harvest energy during the WPT phase and then transmit information during the wireless information transfer (WIT) phase. As for cognitive WPCNs (CWPCNs), in which the SU or the PU is wireless powered, one has to optimize the time and power allocation for maximizing the SU performance while also restricting the interference to the PU or improving the PU performance [10]–[12].

For CWPCNs, by allowing a portion of time exclusively for the wireless powered SU to harvest energy from RF signals, the SU can then act as a friendly jammer to improve the PU security performance and gain its own transmission opportunity simultaneously during the remaining portion of time. By doing so, since the PU security performance is improved, the PU has incentive to let the SU harvest energy and transmit information on its licensed spectrum band. Such cooperation leads to a win-win situation for both the CR networks and the PU networks. This motivates the work in this paper.

In this paper, we propose a new protocol for a CWPCN to cooperate with a PU network who faces security threats

Ding Xu is with the Wireless Communication Key Lab of Jiangsu Province, Nanjing University of Posts and Telecommunications, Nanjing 210003, China (e-mail: xuding@ieee.org). Qun Li is with the Jiangsu Key Lab of Big Data Security and Intelligent Processing, Nanjing University of Posts and Telecommunications, Nanjing 210046, China. (e-mail: liqun@njupt.edu.cn).

from EAVs. In the proposed protocol, the SUs first harvest energy from the power signals transmitted by a half-duplex cognitive hybrid access point (CHAP) during the WPT phase, and then by using the harvested energy, at most one SU is scheduled to transmit information to the CHAP while the other SUs are sending artificial noise to interfere with the EAVs to improve the PU security performance during the WIT phase. The CHAP is responsible for transmitting energy wirelessly to the SU in the downlink and receiving information from the SU in the uplink. For those SUs far from the CHAP and very close to the PU transmitter, the PU transmitter can also decide to broadcast energy to these SUs during the WPT phase. It is assumed that the PU has no knowledge of the channel state information (CSI) related to the EAVs and thus transmits at a constant secrecy rate. Therefore, the secrecy outage probability is adopted as a performance metric for the PU. To encourage the PU to cooperate with the SU, we introduce a new constraint named as the PU secrecy constraint, which requires that the PU secrecy outage probability with the cooperation from the SU shall be lowered to a desired target threshold compared to the case without the cooperation.

Our goal is to maximize the SU ergodic rate by optimizing SU scheduling, power and time allocation under the PU secrecy constraint. Compared to the work such as [5], [13] that considered the SUs are powered by conventional energy sources, the investigated problem in this paper is more intractable due to the fact that the transmit power of the SU depends on the time allocated for the WPT phase and the power allocation of the PU, which makes the optimization variables coupled together and hard to be optimized. We first consider the case when perfect CSI is available at the SUs and the EAVs are not collusive. In this case, by deriving some important properties of the optimal solution, the optimization problem is simplified and an algorithm based on the dual optimization method and the block coordinate descent (BCD) method is proposed to address the problem. We then extend the work to consider the case when the EAVs are collusive. Besides, aiming at minimizing the PU secrecy outage probability, more PU favorable greedy algorithms are also proposed. Furthermore, we also consider the case when the CSI related to the EAVs is unknown to the SUs and propose an efficient algorithm to improve the PU security performance. Extensive simulation results confirm the effectiveness of our proposed CWPCN protocol and corresponding resource allocation algorithms.

It is noted that using wireless powered SUs to improve the PU security performance is appealing due to the following considerations: 1) Since the SU is wirelessly powered, there is no need to replace the exhausted battery and thus the lifetime of the SU is theoretically infinite; 2) Since no battery is needed for the SU, the size of the SU equipment can be small. Thus, the deployment of the SU is flexible and will not get attention by the EAVs. It is also noted that the shortcoming of low available transmit power of the SU can be overcame by deploying the SUs close to the CHAP or the PU transmitter, while the effect of interfering to the EAVs can be improved by increasing the number of SUs and deploying the SUs close to the EAVs.

The main contributions of this paper are summarized as follows:

- We propose a new cooperative protocol for the CWPCN to coexist with the PU network and improve the PU security performance, where the wireless powered SUs first harvest energy, and then at most one SU is scheduled to transmit information while the remaining SUs are sending artificial noise to interfere with the EAVs to improve the PU security performance.
- With the proposed protocol, we first consider the case when perfect CSI is available at the SUs and the EAVs are not collusive, and maximize the SU ergodic rate under the PU secrecy constraint by jointly optimizing SU scheduling, power and time allocation. The optimization problem is shown to be highly non-convex, and we propose a suboptimal algorithm by applying the dual optimization method and the BCD method.
- We then extend the work to consider the case when the EAVs are collusive. Besides, more PU favorable greedy algorithms are also proposed aiming at minimizing the PU secrecy outage probability. For the more practical case when the CSI related to the EAVs is unknown to the SUs, we furthermore propose an efficient algorithm for the SUs to improve the PU security performance.
- Extensive simulation results reveal interesting phenomenons such as increasing the number of EAVs actually improves the SU performance while it degrades the PU security performance, and compared to non-collusive EAVs, collusive EAVs are actually more beneficial to the SUs while they are more harmful to the PU. Besides, it is shown that with unknown EAVs' CSI, the PU security performance can still be improved significantly at the sacrifice of the SU performance.

The rest of this paper is organized as follows. The work related to this paper is described in Section II. Section III presents the system model and formulates the optimization problem. Section IV proposes the resource allocation algorithm for the case of perfect CSI with non-collusive EAVs. Section V extends the work to consider the case of collusive EAVs, proposes greedy algorithms to minimize the PU secrecy outage probability, and furthermore considers the more practical unknown EAVs' CSI case. Section VI verifies the proposed algorithms with extensive simulation results. Finally, Section VII concludes the paper.

## II. RELATED WORK

The PU security performance with the cooperation of the SU was investigated in [5], [13]–[19]. Specifically, in [5], the resource allocation problem to maximize the SU ergodic rate under the condition that the PU secrecy outage probability does not deteriorate due to the existence of the SU was investigated. In [13], besides the PU, the SU was also considered to be threatened by the EAVs, and the SU ergodic secrecy rate was maximized with imperfect CSI under the PU secrecy outage constraint. In [14], the SU secrecy rate was maximized by optimizing power allocation, time allocation and relay selection subject to the minimum PU secrecy rate constraint for both perfect and imperfect CSI. In [15], the SU was proposed to interfere with the EAV and the PU secrecy outage probability was derived in closed-form with fixed values of the SU configuration parameters such as transmit power. In [16], the transmit power of the SU was minimized under different SINR constraints at the EAV, the PU and the SU. In [17], the PU secrecy rate and the SU secrecy rate were optimized based on Stackelberg game. In [18], the power allocation problem to maximize the SU rate under the zero PU secrecy rate loss requirement was investigated. In [19], the close-form expressions of ergodic PU secrecy rate were analyzed with different SU selection schemes. Note that, unlike this paper, the above references [5], [13]–[19] all assumed that the SU is equipped with constant energy supplies.

The physical-layer security in WPCNs has been investigated in [20]–[26]. In [20], the source node was assumed to harvest energy from the base station and use these energy for interfering with the EAVs, and the secrecy throughput was maximized under the secrecy constraint and the transmit power constraint. In [21], a simple communication protocol for secure communications with the help from a wireless powered jammer was proposed and the secrecy throughput was maximized. In [22], a two-phase protocol, where the destination node was assumed to harvest energy from the source node in the first phase and then use these energy to interfere with the EAVs in the second phase, was proposed and the secrecy energy efficiency was maximize by optimizing the time allocation and the power allocation. In [23], by considering both the minimum secrecy rate requirement and the minimum harvested energy requirement, the secrecy energy efficiency was maximized by optimizing power allocation and power splitting ratio. In [24], a WPCN with a wireless powered jammer was considered, and the time allocation was optimized to maximize the secrecy rate and minimize the secrecy outage probability. In [25], a WPCN with a wireless powered jammer and an energy receiver who is a potential EAV was considered, and the power allocation at the source node and the jammer to maximize the secrecy rate was optimized. In [26], a multicarrier WPCN with a wireless powered jammer was considered, and the secrecy rate was maximized by jointly optimizing time allocation and power allocation at the source node and the jammer over subcarriers. Note that, compared to these works on security in WPCNs in [20]–[26], our problem is different due to the need to improve the PU security performance as well as providing transmission opportunities for the SU.

As for the physical-layer security in CWPCNs, some recent work focused on investigating secure communications for the SU, such as [27]–[32]. In [27], the secrecy outage probability of the SU was derived in closed-form under the interference power constraint at the PU. In [28], multiple design objectives for the SU were considered under the secrecy performance guarantee at the SU and the interference power constraint at the PU. In [29], the PU was assumed to be a potential EAV for the SU and the SU transmit power was minimized under the minimum SU secrecy rate and the minimum harvested energy constraints. In [30], under the interference power constraint at the PU and assuming that the secondary receiver is full-duplex and sends jamming signals to the EAV, the probability of strictly positive secrecy rate at the SU was investigated. In [31], the SU secrecy rate was improved by letting a wireless powered jammer to degrade the performance of the EAV. In [32], a SU scheduling scheme was proposed for a CWPCN with the interference power constraint at the PU, and its closed-form expressions of outage probability and intercept probability were derived over Rayleigh fading channels. Compared to these related works in [27]–[32], our work is different in that secure communications for the PU is our concern.

The work in [33]–[37] considered PU security in CWPCNs. In [33]–[35], the SU was assumed to harvest energy from the environment or the PU signals and then relay PU messages to enhance PU secrecy performance. Note that [33]–[35] considered only one pair of SUs and the proposed schemes can improve PU secrecy performance only if the SU is far from the EAV. In [36], the wireless powered SU was considered as a potential EAV for the PU and the SU rate was maximized under the SINR constraints at the PU and the SU. In [37], the secrecy guard zones were set by the primary network to protect itself from eavesdropping from the wireless powered secondary network. It is noted that the SU was considered as a potential EAV in [36], [37], while this paper treats the SU as a friendly jammer for the PU. To our best knowledge, no work in existing literature has considered to use wireless powered SUs to act as friendly jammers to improve the PU security performance and gain their own transmission opportunities at the same time.

## III. SYSTEM MODEL

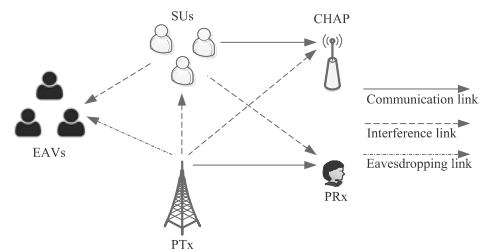

Figure 1. System model.

We consider an uplink CWPCN with $K$ SUs and one CHAP sharing the narrow spectrum band with one pair of PUs (a primary transmitter (PTx) and a primary receiver (PRx)) who faces security threats from $N$ EAVs, as shown in Fig. 1. Let $\mathbb{K}$ and $\mathbb{N}$ denote the sets of SUs and EAVs, respectively. It is assumed that the SUs are powered only by the energy harvested from RF signals while the rest nodes are powered by constant energy sources. All the channels involved are assumed to be block fading, i.e., all the channel power gains are constant within a transmission block and may change independently from one block to another. The channel power gains between SU $k \in \mathbb{K}$ and the CHAP, from SU $k \in \mathbb{K}$ to the PRx, from SU $k \in \mathbb{K}$ to EAV $n \in \mathbb{N}$, from the PTx to the PRx, from the PTx to SU $k \in \mathbb{K}$, from the PTx to the CHAP, and from the PTx to EAV $n \in \mathbb{N}$ are denoted by $h_{ss}^k[\nu]$, $h_{sp}^k[\nu]$, $h_{se}^{k,n}[\nu]$, $h_{pp}[\nu]$, $h_{pst}^k[\nu]$, $h_{psr}[\nu]$, and $h_{pe}^n[\nu]$, respectively, where $\nu$ denotes the joint fading state for all the channels involved. It is assumed that the PU has no knowledge on the CSI of the links related to the EAVs and thus transmits at a constant secrecy rate $R$. As for the SUs' knowledge on the CSI, two cases are considered. In the first case, the CSI is assumed to be perfectly available at the SUs. In the second case, the CSI of the links involving the EAVs is assumed to be unavailable while the other CSI is assumed to be perfectly known at the SUs[1]. It is noted that although perfect CSI is hard to be realized in practice, it can be served as an upper bound or a benchmark for imperfect CSI. Investigating the perfect CSI case can let us know the potential benefits of such cooperation between the SU and the PU and guide the design of the imperfect CSI case.

For each transmission block, without loss of generality, the time is assumed to be unit. The unit time is assumed to be divided into two phases: the WPT phase and the WIT phase. In the first WPT phase with time duration $\tau_0[\nu]$, the CHAP and the PTx broadcast RF energy to the SUs, and meanwhile the SUs harvest energy from these RF signals and then store the energy in supercapacitors. Compared to rechargeable batteries, supercapacitors have smaller form-factor, faster charging cycle and longer life [39]. However, energy leakage for a supercapacitor is very large [40] and thus the energy might not be able to be stored long enough for the next transmission block. For simplicity, similar to [41], we assume that the harvested energy in current transmission block cannot be used for future transmission blocks. In the second WIT phase with time duration $\tau_1[\nu]$, the PTx transmits information to the PRx, and meanwhile at most one SU is scheduled to transmit information to the CHAP and the other SUs are sending artificial noise to interfere with the EAVs.

Supposing that the cooperation from the SU is not available and all the time within each transmission block are used by the PU for information transmission with maximum transmit power $P_{max}$, then the achievable rate at the PU and the EAV $n$ at fading state $\nu$ can be written as [5]

$$R_{wos,p}[\nu] = \log_2\left(1 + \frac{P_{max}h_{pp}[\nu]}{\sigma^2}\right), \quad (1)$$

and

$$R_{wos,e}^n[\nu] = \log_2\left(1 + \frac{P_{max}h_{pe}^n[\nu]}{\sigma^2}\right), \quad (2)$$

respectively, where $\sigma^2$ is the noise power. Then, if the EAVs are non-collusive, the maximum secrecy rate of the PU is [5]

$$C_{wos}[\nu] = \left(R_{wos,p}[\nu] - \max_{n \in \mathbb{N}} R_{wos,e}^n[\nu]\right)^+, \quad (3)$$

where $(.)^+$ denotes $\max(.,0)$. If the EAVs are collusive and use maximum ratio combining, the achievable rate at the EAVs is given as [42]

$$R_{wos,e}[\nu] = \log_2\left(1 + \frac{P_{max}\sum_{n \in \mathbb{N}} h_{pe}^n[\nu]}{\sigma^2}\right), \quad (4)$$

---

[1] Note that the effects of the imperfect CSI of the links not involving EAVs (such as the links between the SUs and the CHAP, and the links between the SUs and the PU) have been well researched in existing literature such as [38]. Thus, we focus on our main contributions of this paper and consider the CSI of the links related to the EAVs is unknown.

and the maximum secrecy rate of the PU is given as [42]

$$C_{wos}[\nu] = \left(R_{wos,p}[\nu] - R_{wos,e}[\nu]\right)^+. \quad (5)$$

Since the PU transmits at a constant secrecy rate $R$, the PU secrecy outage probability is given by $\varepsilon_p = \Pr\{C_{wos}[\nu] < R\}$. Note that, $\varepsilon_p$ is a known value if $P_{max}$, $R$ and the distribution of $h_{pp}[\nu]$ and $h_{pe}^n[\nu]$ are given.

Now, supposing that the SU cooperates with the PU. In this case, the CHAP transmits RF energy with constant power $Q$ during WPT. In addition, the PTx transmits RF energy with power $p_0[\nu]$ during WPT and transmits information with power $p_1[\nu]$ during WIT at fading state $\nu$. Note that $p_0[\nu]$ can be equal to zero and this means the PTx will not transfer energy to the SUs. The energy harvested by SU $k \in \mathbb{K}$ during WPT is $\lambda(h_{pst}^k[\nu]p_0[\nu] + h_{ss}^k[\nu]Q)\tau_0[\nu]$, where $\lambda$ ($0 < \lambda < 1$) is the energy harvesting efficiency[2]. The energy consumed by SU $k \in \mathbb{K}$ during WIT cannot be higher than that harvested during WPT. Let $p_s^k[\nu]$ and $q_s^k[\nu]$ denote the transmit powers of SU $k \in \mathbb{K}$ for information transmission and jamming during WIT at fading state $\nu$, respectively. Then, we have $(p_s^k[\nu] + q_s^k[\nu])\tau_1[\nu] \leq \lambda(h_{pst}^k[\nu]p_0[\nu] + h_{ss}^k[\nu]Q)\tau_0[\nu]$. Since each SU is scheduled either for information transmission or jamming and at most one SU is scheduled for information transmission, we have $p_s^k[\nu]q_s^k[\nu] = 0$ for all $k \in \mathbb{K}$ and $p_s^k[\nu]p_s^{k'}[\nu] = 0$, $\forall k \neq k'$. Similar to the existing work such as [14], [15], we assume that the PU and the EAVs treat the information signals from the SU as interference. We also assume that the artificial noise sent from the SUs can be canceled at the PU and the CHAP but cannot be canceled at the EAVs. Such assumption can be practically realized by the method proposed in [44]: the SUs, the CHAP and the PU pre-store a set of Gaussian distributed random sequences used for artificial noise with the indices of the sequences treated as the keys. The SU randomly select a sequence and send its key secretly to the CHAP and the PU using the phase-shift modulation-based method proposed in [45]. Since the set of random sequences is unknown to the EAVs, the EAVs cannot know the sequence used by the SU even if the key is intercepted by the EAVs.

Then, with SUs' cooperation, the achievable rate at the PU and the EAV $n$ at fading state $\nu$ can be written as

$$R_{ws,p}[\nu] = \tau_1[\nu]\log_2\left(1 + \frac{p_1[\nu]h_{pp}[\nu]}{\sigma^2 + \sum_{k \in \mathbb{K}} p_s^k[\nu]h_{sp}^k[\nu]}\right), \quad (6)$$

and

$$R_{ws,e}^n[\nu] = \tau_1[\nu]\log_2\left(1 + \frac{p_1[\nu]h_{pe}^n[\nu]}{\sigma^2 + \sum_{k \in \mathbb{K}}(p_s^k[\nu] + q_s^k[\nu])h_{se}^{k,n}[\nu]}\right), \quad (7)$$

respectively. Supposing that the EAVs are non-collusive, the maximum secrecy rate of the PU with SUs' cooperation at fading state $\nu$ is

$$C_{ws}[\nu] = \left(R_{ws,p}[\nu] - \max_{n \in \mathbb{N}} R_{ws,e}^n[\nu]\right)^+. \quad (8)$$

---

[2] Note that the energy harvesting efficiency achieved by commercial products can be as high as 95% [43].

Supposing that the EAVs are collusive and use maximum ratio combining, the achievable rate at the EAVs is given as

$$R_{ws,e}[\nu] = \tau_1[\nu]$$
$$\times \log_2\left(1 + \sum_{n\in\mathbb{N}} \frac{p_1[\nu]h_{pe}^n[\nu]}{\sigma^2 + \sum_{k\in\mathbb{K}}(p_s^k[\nu]+q_s^k[\nu])h_{se}^{k,n}[\nu]}\right), \quad (9)$$

and the maximum secrecy rate of the PU is given as

$$C_{ws}[\nu] = \left(R_{ws,p}[\nu] - R_{ws,e}[\nu]\right)^+. \quad (10)$$

It is noted that scheduling the SU near the PRx for information transmission will cause severe interference to the PRx. Thus through proper SU scheduling, the SU who causes minor interference to the PRx shall be scheduled for information transmission. If such SU does not exist, no SU shall be scheduled for information transmission.

The PU secrecy outage probability with the cooperation from the SU is written as $\varepsilon_{ps} = \Pr\{C_{ws}[\nu] < R\}$. To encourage the PU to cooperate with the SU, the PU secrecy outage probability $\varepsilon_{ps}$ is required to be lower than that without the SU's cooperation, i.e.,

$$\varepsilon_{ps} \leq \varepsilon_p - \triangle\varepsilon_p = \varepsilon_0, \quad (11)$$

where $\triangle\varepsilon_p$ ($\triangle\varepsilon_p \geq 0$) denotes the required PU secrecy outage probability reduction[3]. We denote the above constraint as the PU secrecy constraint.

We aim to optimize SU scheduling, power and time allocation to maximize the SU ergodic rate as[4] (P1)

$$\max_{\{p_s^k\},\{q_s^k\},p_0,p_1,\tau_0,\tau_1} E\left\{\tau_1 \sum_{k\in\mathbb{K}} \log_2\left(1 + \frac{p_s^k h_{ss}^k}{\sigma^2 + p_1 h_{psr}}\right)\right\} \quad (12)$$

s.t. $p_0\tau_0 + p_1\tau_1 \leq P_{max},$ (13)

$\tau_0 + \tau_1 \leq 1,$ (14)

$(p_s^k + q_s^k)\tau_1 \leq \lambda(h_{pst}^k p_0 + h_{ss}^k Q)\tau_0, \forall k \in \mathbb{K}$ (15)

$p_s^k q_s^k = 0, \forall k \in \mathbb{K}$ (16)

$p_s^k p_s^{k'} = 0, \forall k \neq k'$ (17)

$\varepsilon_{ps} \leq \varepsilon_0,$ (18)

$p_s^k \geq 0, q_s^k \geq 0, \forall k \in \mathbb{K}, p_0 \geq 0, p_1 \geq 0, \tau_0 \geq 0, \tau_1 \geq 0.$ (19)

In P1, the objective function in (12) is the SU ergodic rate defined as the rate averaged over all transmission fading blocks. Here, we assume that the SU carries delay-tolerant services. Such assumption is reasonable as the SU transmission is opportunistic and thus stringent transmission delay requirement for the SU is improper. Since ergodic rate can be considered as the average rate under no transmission delay requirement, we adopt ergodic rate as our performance metric for the SU. The constraint in (13) is used to restrict the transmit powers of the PU in the WPT and WIT phases such that the average transmit power over the two phases is below $P_{max}$. The constraint in (13) is also used for a fair comparison with the case without SU's cooperation. Since unit transmission block is assumed, the constraint in (13) can also be interpreted as a constraint to restrict the total energy consumed by the PU in the two phases. The constraint in (14) indicates that the total time length of the two phases must be no larger than 1. The constraint in (15) indicates that the energy consumed by the SU during WIT must be no larger than the energy harvested during WPT. The constraint in (16) indicates that each SU is scheduled either for information transmission or jamming. The constraint in (17) indicates that at most one SU is scheduled for information transmission. The constraint in (18) is the PU secrecy constraint used to encourage the PU to cooperate with the SU such that the PU secrecy performance in terms of secrecy outage probability can be improved by the cooperation from the SU. It is noted that the value of $\varepsilon_0$ should be not too small (i.e., $\triangle\varepsilon_p$ should be not too large), otherwise P1 will be infeasible. Here, we assume that the value of $\triangle\varepsilon_p$ is small enough such that P1 is feasible[5].

## IV. RESOURCE ALLOCATION ALGORITHM WITH PERFECT CSI AND NON-COLLUSIVE EAVS

This section considers the case of perfect CSI and non-collusive EAVs for P1. To our best knowledge, P1 is a highly non-linear non-convex problem that cannot be solved directly with standard method. To solve P1, we first derive some properties of the optimal solution to P1, and then we simplify P1 using these properties.

*Proposition 1*: The optimal solution to P1 must satisfy the constraint in (13) with strict equality, i.e., $p_0\tau_0 + p_1\tau_1 = P_{max}$.

*Proof*: Please refer to Appendix A. □

*Proposition 2*: The optimal solution to P1 can always satisfy the constraint in (14) with strict equality, i.e., $\tau_0 + \tau_1 = 1$.

*Proof*: Please refer to Appendix B. □

Based on Propositions 1-2, we have the following relationships between the optimization variables for the optimal solution to P1 as

$$p_0 = \frac{P_{max} - p_1\tau_1}{1-\tau_1}, \tau_0 = 1 - \tau_1. \quad (20)$$

Inserting the above expressions into P1, P1 is reduced to the

---

[3]In this paper, we assume that the PU always has data to transmit/receive. In such a case, the PU can cooperate with the SU to improve its own secrecy performance as enforced by the constraint in (11). For the case when the PU has no data to transmit/receive, the SU can occupy the spectrum freely or being charged with a price. The interested readers may refer to [46] and references therein for the work on pricing-based methods dealing with the case when the PU is willing to lease its unused spectrum to the SU.

[4]Hereafter, we omit the index $\nu$ for the fading state wherever it is clear from the context.

[5]It is noted that the greedy algorithms proposed in Section V-B can be used to check the feasibility of P1. As long as $\eta$ in the greedy algorithm is very large, the obtained PU secrecy outage probability is practically the smallest PU secrecy outage probability that can be achieved by cooperation from the SU. Thus, if the PU secrecy outage probability obtained from the greedy algorithm is larger than $\varepsilon_0$, then P1 is infeasible, and otherwise, P1 is feasible.

following problem as (P2)

$$\max_{\{p_s^k\},\{q_s^k\},p_1,\tau_1} E\left\{\tau_1 \sum_{k\in\mathbb{K}} \log_2\left(1+\frac{p_s^k h_{ss}^k}{\sigma^2+p_1 h_{psr}}\right)\right\} \quad (21)$$

$$\text{s.t. } 0 \leq p_1\tau_1 \leq P_{max}, \quad (22)$$

$$0 \leq \tau_1 \leq 1, \quad (23)$$

$$(p_s^k+q_s^k)\tau_1 \leq \lambda(h_{pst}^k P_{max} - h_{pst}^k p_1\tau_1 + h_{ss}^k Q(1-\tau_1)), \forall k \in \mathbb{K}, \quad (24)$$

$$p_s^k \geq 0, q_s^k \geq 0, \forall k \in \mathbb{K}, p_1 \geq 0, \quad (25)$$

and constraints (16)-(18).

To solve P2, we introduce a virtual SU indexed by $k=0$ and schedule it for information transmission if all the actual $K$ SUs are scheduled for jamming. The channel power gains related to this virtual SU are set to zero. In what follows, we assume that SU 0 is included in the SU set $\mathbb{K}$. The following indicator function is introduced as

$$\chi(\{p_s^k\},\{q_s^k\},p_1,\tau_1) = \begin{cases} 1, & C_{ws} < R, \\ 0, & C_{ws} \geq R. \end{cases} \quad (26)$$

Then, the constraint (18) can be rewritten as

$$E\{\chi(\{p_s^k\},\{q_s^k\},p_1,\tau_1)\} \leq \varepsilon_0. \quad (27)$$

It is observed that P2 is still highly non-convex and hard to be solved with standard methods. In the following, we solve P2 with the dual optimization method [47] and the block coordinate descent (BCD) method [48].

It is seen that the optimization variables for all the channel fading states are coupled in constraint (27). Thus, it makes sense to adopt dual decomposition by forming the Lagrangian function to relax the coupling constraint (27) as given by [47]

$$L(\{p_s^k\},\{q_s^k\},p_1,\tau_1,\eta)$$
$$= E\left\{\tau_1 \sum_{k\in\mathbb{K}} \log_2\left(1+\frac{p_s^k h_{ss}^k}{\sigma^2+p_1 h_{psr}}\right)\right\}$$
$$- \eta(E\{\chi(\{p_s^k\},\{q_s^k\},p_1,\tau_1)\} - \varepsilon_0), \quad (28)$$

where $\eta$ is the dual variable associated with the constraint (27). Then, the Lagrange dual function is written as [47]

$$g(\eta) = \max_{\{p_s^k\},\{q_s^k\},p_1,\tau_1} L(\{p_s^k\},\{q_s^k\},p_1,\tau_1,\eta) \quad (29)$$
$$\text{s.t. constraints } (16),(17),(22)-(25).$$

Using (28), we can rewrite the above problem as $g(\eta) = E\{g'(\eta)\} + \eta\varepsilon_0$, where

$$g'(\eta) = \max_{\{p_s^k\},\{q_s^k\},p_1,\tau_1} \tau_1 \sum_{k\in\mathbb{K}} \log_2\left(1+\frac{p_s^k h_{ss}^k}{\sigma^2+p_1 h_{psr}}\right)$$
$$- \eta\chi(\{p_s^k\},\{q_s^k\},p_1,\tau_1) \quad (30)$$
$$\text{s.t. constraints } (16),(17),(22)-(25).$$

Thus, through the dual decomposition, P2 is separated into two levels of optimization. At the lower level, we have the subproblems, one for each channel fading state as given by the problem in (30). At the higher level, we have the master dual problem written as $\min_{\eta\geq 0} g(\eta)$. The optimal solution obtained from the two levels of optimization is the optimal solution of P2 if the duality gap is zero [47], otherwise, the solution is a feasible solution of P2. Luckily, it has been shown in [49] that the time-sharing condition is satisfied if the channel power gain distributions are continuous for a problem similar to P2, and the duality gap is zero by satisfying the time-sharing condition. Therefore, adopting the dual optimization is appropriate for P2. Since the master dual problem can be easily solved by updating $\eta$ via the subgradient method until convergence [47], we focus on solving the problem in (30) in what follows.

Since the problem in (30) is nonconvex, the optimal solution is computationally difficult to derive. Instead, we adopt the BCD method to obtain a suboptimal solution by iteratively optimizing $\tau_1$ with given $\{p_s^k\},\{q_s^k\},p_1$, optimizing $p_1$ with given $\{p_s^k\},\{q_s^k\},\tau_1$, and optimizing $\{p_s^k\},\{q_s^k\}$ with given $p_1,\tau_1$ until convergence. Firstly, with given $\{p_s^k\},\{q_s^k\},p_1$, the value of $\tau_1$ is optimized as

$$\max_{\tau_1} \tau_1 \sum_{k\in\mathbb{K}} \log_2\left(1+\frac{p_s^k h_{ss}^k}{\sigma^2+p_1 h_{psr}}\right)$$
$$- \eta\chi(\{p_s^k\},\{q_s^k\},p_1,\tau_1) \quad (31)$$

$$\text{s.t. } 0 \leq \tau_1 \leq \min\left(\min\left(\frac{P_{max}}{p_1},1\right),\right.$$
$$\left.\min_k\left(\frac{\lambda h_{pst}^k P_{max} + \lambda h_{ss}^k Q}{p_s^k+q_s^k+\lambda h_{pst}^k p_1+\lambda h_{ss}^k Q}\right)\right). \quad (32)$$

Since $C_{ws}$ increases as $\tau_1$ increases if $C_{ws}$ is nonzero or keeps unchanged as $\tau_1$ increases if $C_{ws}$ is zero, the indicator function $\chi(\{p_s^k\},\{q_s^k\},p_1,\tau_1)$ may decrease from 1 to 0 or remain unchanged as $\tau_1$ increases. Thus, the objective function in (31) is maximized at the maximum allowable $\tau_1$ given by

$$\tau_1 = \min\left(\min\left(\frac{P_{max}}{p_1},1\right),\right.$$
$$\left.\min_k\left(\frac{\lambda h_{pst}^k P_{max} + \lambda h_{ss}^k Q}{p_s^k+q_s^k+\lambda h_{pst}^k p_1+\lambda h_{ss}^k Q}\right)\right). \quad (33)$$

Next, with given $\{p_s^k\},\{q_s^k\},\tau_1$, the value of $p_1$ is optimized as

$$\max_{p_1} \tau_1 \sum_{k\in\mathbb{K}} \log_2\left(1+\frac{p_s^k h_{ss}^k}{\sigma^2+p_1 h_{psr}}\right) - \eta\chi(\{p_s^k\},\{q_s^k\},p_1,\tau_1) \quad (34)$$

$$\text{s.t. } 0 \leq p_1 \leq \bar{p}_1, \quad (35)$$

where $\bar{p}_1 = \min(P_{max}/\tau_1, \min_k(\lambda h_{pst}^k P_{max} + \lambda h_{ss}^k Q - (\lambda h_{ss}^k Q + p_s^k + q_s^k)\tau_1)/\lambda h_{pst}^k \tau_1)$. Let $f_1(p_1) = \tau_1 \sum_{k\in\mathbb{K}} \log_2\left(1+\frac{p_s^k h_{ss}^k}{\sigma^2+p_1 h_{psr}}\right)$. It is easily seen that $f_1(p_1)$ is a monotonically decreasing function of $p_1$. It is noted that the objective function in (34) now becomes $f_1(p_1) - \eta\chi(\{p_s^k\},\{q_s^k\},p_1,\tau_1)$. From (26), if $C_{ws} \geq R$, we have $\chi(\{p_s^k\},\{q_s^k\},p_1,\tau_1) = 0$. From (6)-(8), we can reformulate the inequality $C_{ws} \geq R$ as a set of inequalities

given by $C_{ws}^n \geq R, n \in \mathbb{N}$, where

$$C_{ws}^n = \tau_1 \left( \log_2 \left( 1 + \frac{p_1 h_{pp}}{\sigma^2 + \sum_{k \in \mathbb{K}} p_s^k h_{sp}^k} \right) \right.$$
$$\left. - \log_2 \left( 1 + \frac{p_1 h_{pe}^n}{\sigma^2 + \sum_{k \in \mathbb{K}} (p_s^k + q_s^k) h_{se}^{k,n}} \right) \right)^+, n \in \mathbb{N}. \quad (36)$$

The following indicator function is then introduced as

$$\chi_n(\{p_s^k\}, \{q_s^k\}, p_1, \tau_1) = \begin{cases} 1, & C_{ws}^n < R, \\ 0, & C_{ws}^n \geq R. \end{cases} \quad (37)$$

Based on the above reformulation, we can rewrite $\chi(\{p_s^k\}, \{q_s^k\}, p_1, \tau_1)$ as $\chi(\{p_s^k\}, \{q_s^k\}, p_1, \tau_1) = \max_{n \in \mathbb{N}} \chi_n(\{p_s^k\}, \{q_s^k\}, p_1, \tau_1)$. Then, the problem in (34) is reformulated as

$$\min_{n \in \mathbb{N}} \max_{p_1} f_1(p_1) - \eta \chi_n(\{p_s^k\}, \{q_s^k\}, p_1, \tau_1) \quad (38)$$
$$\text{s.t. } 0 \leq p_1 \leq \bar{p}_1. \quad (39)$$

Based on the above reformulation of the problem in (34), its solution can be obtained by first deriving $N$ optimal $p_1$'s, one for each of the inner maximization problems in (38), and then picking the one that minimizes the objective function in (38). The following theorem presents the optimal solution to the problem in (38).

*Theorem 1*: The optimal solution to the problem in (38) is given by $p_1 = p_1^{n^*}$, where $n^* = \arg\min_{n \in \mathbb{N}} f_1(p_1^n) - \eta \chi_n(\{p_s^k\}, \{q_s^k\}, p_1^n, \tau_1)$ and $p_1^n$ is given by

$$p_1^n = \begin{cases} \frac{2^{\frac{R}{\tau_1}} - 1}{W_n}, & W_n > 0, \frac{2^{\frac{R}{\tau_1}} - 1}{W_n} \leq \bar{p}_1, \\ & f_1\left(\frac{2^{\frac{R}{\tau_1}} - 1}{W_n}\right) \geq f_1(0) - \eta, \\ 0, & \text{otherwise,} \end{cases} \quad (40)$$

and $W_n = \frac{h_{pp}}{\sigma^2 + \sum_{k \in \mathbb{K}} p_s^k h_{sp}^k} - \frac{2^{\frac{R}{\tau_1}} h_{pe}^n}{\sigma^2 + \sum_{k \in \mathbb{K}} (p_s^k + q_s^k) h_{se}^{k,n}}$.

*Proof*: Please refer to Appendix C. $\square$

Then, with given $p_1, \tau_1$, the value of $\{p_s^k\}, \{q_s^k\}$ is optimized as

$$\max_{\{p_s^k\}, \{q_s^k\}} \tau_1 \sum_{k \in \mathbb{K}} \log_2 \left( 1 + \frac{p_s^k h_{ss}^k}{\sigma^2 + p_1 h_{psr}} \right)$$
$$- \eta \chi(\{p_s^k\}, \{q_s^k\}, p_1, \tau_1) \quad (41)$$
$$\text{s.t. } p_s^k q_s^k = 0, \forall k \in \mathbb{K} \quad (42)$$
$$p_s^k p_s^{k'} = 0, \forall k \neq k' \quad (43)$$
$$p_s^k + q_s^k \leq \frac{\lambda(h_{pst}^k P_{max} + h_{ss}^k Q)}{\tau_1}$$
$$- \lambda(h_{pst}^k p_1 + h_{ss}^k Q), \forall k \in \mathbb{K}, \quad (44)$$
$$p_s^k \geq 0, q_s^k \geq 0, \forall k \in \mathbb{K}. \quad (45)$$

Since only one SU can be scheduled for information transmission, i.e., only one $p_s^k$ can be nonzero, we can solve the problem in (41) by first deriving $K + 1$ optimal power allocations, one for each of the total $K + 1$ SUs, and then selecting the one that maximizes the objective function in (41). Assuming that SU $\tilde{k} \in \mathbb{K}$ is scheduled for information transmission, i.e., $q_s^{\tilde{k}} = 0, p_s^k = 0, \forall k \neq \tilde{k}$, then the secrecy rate of the PU $C_{ws}$ becomes

$$C_{ws}^{\tilde{k}} = \tau_1 \left( \log_2 \left( 1 + \frac{p_1 h_{pp}}{\sigma^2 + p_s^{\tilde{k}} h_{sp}^{\tilde{k}}} \right) - \max_{n \in \mathbb{N}} \log_2 \left( 1 \right. \right.$$
$$\left. \left. + \frac{p_1 h_{pe}^n}{\sigma^2 + \sum_{k \neq \tilde{k}} q_s^k h_{se}^{k,n} + p_s^{\tilde{k}} h_{se}^{\tilde{k},n}} \right) \right)^+, \quad (46)$$

and constraint (44) is rewritten as

$$0 \leq p_s^{\tilde{k}} \leq \bar{p}_s^{\tilde{k}}, \quad (47)$$
$$0 \leq q_s^k \leq \bar{q}_s^k, \forall k \neq \tilde{k}, \quad (48)$$

where

$$\bar{p}_s^{\tilde{k}} = \frac{\lambda(h_{pst}^{\tilde{k}} P_{max} + h_{ss}^{\tilde{k}} Q)}{\tau_1} - \lambda(h_{pst}^{\tilde{k}} p_1 + h_{ss}^{\tilde{k}} Q), \quad (49)$$
$$\bar{q}_s^k = \frac{\lambda(h_{pst}^k P_{max} + h_{ss}^k Q)}{\tau_1} - \lambda(h_{pst}^k p_1 + h_{ss}^k Q), \forall k \neq \tilde{k}. \quad (50)$$

Clearly, $C_{ws}^{\tilde{k}}$ in (46) is a non-decreasing function of $q_s^k$. Thus, the transmit power of the SU scheduled for jamming can be set to its maximum allowable value, i.e., $q_s^k = \bar{q}_s^k, \forall k \neq \tilde{k}$, in order to decrease the PU secrecy outage probability as low as possible. Therefore, assuming that SU $\tilde{k} \in \mathbb{K}$ is scheduled for information transmission, the problem in (41) is simplified as

$$\max_{p_s^{\tilde{k}}} \tau_1 \log_2 \left( 1 + \frac{p_s^{\tilde{k}} h_{ss}^{\tilde{k}}}{\sigma^2 + p_1 h_{psr}} \right) - \eta \chi^{\tilde{k}}(p_s^{\tilde{k}}) \quad (51)$$
$$\text{s.t. } 0 \leq p_s^{\tilde{k}} \leq \bar{p}_s^{\tilde{k}}. \quad (52)$$

where

$$\chi^{\tilde{k}}(p_s^{\tilde{k}}) = \begin{cases} 1, & C_{ws}^{\tilde{k}} < R, \\ 0, & \text{otherwise,} \end{cases} \quad (53)$$

and

$$C_{ws}^{\tilde{k}} = \tau_1 \left( \log_2 \left( 1 + \frac{p_1 h_{pp}}{\sigma^2 + p_s^{\tilde{k}} h_{sp}^{\tilde{k}}} \right) - \max_{n \in \mathbb{N}} \log_2 \left( 1 \right. \right.$$
$$\left. \left. + \frac{p_1 h_{pe}^n}{\sigma^2 + \sum_{k \neq \tilde{k}} \bar{q}_s^k h_{se}^{k,n} + p_s^{\tilde{k}} h_{se}^{\tilde{k},n}} \right) \right)^+. \quad (54)$$

Let $f^{\tilde{k}}(p_s^{\tilde{k}}) = \tau_1 \log_2 \left( 1 + \frac{p_s^{\tilde{k}} h_{ss}^{\tilde{k}}}{\sigma^2 + p_1 h_{psr}} \right)$. It is easily seen that $f^{\tilde{k}}(p_s^{\tilde{k}})$ is a monotonically increasing function of $p_s^{\tilde{k}}$. It is noted that the objective function in (51) now becomes $f^{\tilde{k}}(p_s^{\tilde{k}}) - \eta \chi^{\tilde{k}}(p_s^{\tilde{k}})$. From (54), the inequality $C_{ws}^{\tilde{k}} \geq R$ can be rewritten as a set of inequalities given by $C_{ws}^{\tilde{k},n} \geq R, n \in \mathbb{N}$, where

$$C_{ws}^{\tilde{k},n} = \tau_1 \left( \log_2 \left( 1 + \frac{p_1 h_{pp}}{\sigma^2 + p_s^{\tilde{k}} h_{sp}^{\tilde{k}}} \right) - \log_2 \left( 1 \right. \right.$$
$$\left. \left. + \frac{p_1 h_{pe}^n}{\sigma^2 + \sum_{k \neq \tilde{k}} \bar{q}_s^k h_{se}^{k,n} + p_s^{\tilde{k}} h_{se}^{\tilde{k},n}} \right) \right)^+, n \in \mathbb{N}. \quad (55)$$

Then, we can rewrite $\chi^{\tilde{k}}(p_s^{\tilde{k}})$ as $\chi^{\tilde{k}}(p_s^{\tilde{k}}) = \max_{n \in \mathbb{N}} \chi_n^{\tilde{k}}(p_s^{\tilde{k}})$, where

$$\chi_n^{\tilde{k}}(p_s^{\tilde{k}}) = \begin{cases} 1, & C_{ws}^{\tilde{k},n} < R, \\ 0, & C_{ws}^{\tilde{k},n} \geq R. \end{cases} \quad (56)$$

Then, the problem in (51) can be rewritten as

$$\min_{n \in \mathbb{N}} \max_{p_s^{\tilde{k}}} f^{\tilde{k}}(p_s^{\tilde{k}}) - \eta \chi_n^{\tilde{k}}(p_s^{\tilde{k}}) \qquad (57)$$

$$\text{s.t.} \ 0 \leq p_s^{\tilde{k}} \leq \bar{p}_s^{\tilde{k}}. \qquad (58)$$

Therefore, the problem in (51) can be solved by first solving the inner maximization problem in (57) for each $n \in \mathbb{N}$ and then selecting the one that minimizes the objective function in (57). The following theorem presents the optimal solution to the problem in (57).

*Theorem 2*: The optimal solution to the problem in (57) is given by $p_s^{\tilde{k}} = p_s^{\tilde{k}, n^*}$, where $p_s^{\tilde{k}, n}$ is given by

$$p_s^{\tilde{k}, n} = \begin{cases} -\frac{B_{\tilde{k}, n}}{2A_{\tilde{k}, n}}, & B_{\tilde{k}, n}^2 - 4A_{\tilde{k}, n}C_{\tilde{k}, n} = 0, 0 \leq -\frac{B_{\tilde{k}, n}}{2A_{\tilde{k}, n}} \leq \bar{p}_s^{\tilde{k}}, \\ & f^{\tilde{k}}(-\frac{B_{\tilde{k}, n}}{2A_{\tilde{k}, n}}) \geq f^{\tilde{k}}(\bar{p}_s^{\tilde{k}}) - \eta \\ x_{\tilde{k}, n, 2}, & B_{\tilde{k}, n}^2 - 4A_{\tilde{k}, n}C_{\tilde{k}, n} > 0, 0 \leq x_{\tilde{k}, n, 2} \leq \bar{p}_s^{\tilde{k}}, \\ & f^{\tilde{k}}(x_{\tilde{k}, n, 2}) \geq f^{\tilde{k}}(\bar{p}_s^{\tilde{k}}) - \eta, \\ \bar{p}_s^{\tilde{k}}, & \text{otherwise}, \end{cases}$$
(59)

and

$$n^* = \arg\min_{n \in \mathbb{N}} f^{\tilde{k}}(p_s^{\tilde{k}, n}) - \eta \chi_n^{\tilde{k}}(p_s^{\tilde{k}, n}), \qquad (60)$$

$$x_{\tilde{k}, n, 1} = \frac{-B_{\tilde{k}, n} - \sqrt{B_{\tilde{k}, n}^2 - 4A_{\tilde{k}, n}C_{\tilde{k}, n}}}{2A_{\tilde{k}, n}}, \qquad (61)$$

$$x_{\tilde{k}, n, 2} = \frac{-B_{\tilde{k}, n} + \sqrt{B_{\tilde{k}, n}^2 - 4A_{\tilde{k}, n}C_{\tilde{k}, n}}}{2A_{\tilde{k}, n}}, \qquad (62)$$

$$A_{\tilde{k}, n} = (2^{\frac{R}{\tau_1}} - 1) h_{sp}^{\tilde{k}} h_{se}^{\tilde{k}, n}, \qquad (63)$$

$$B_{\tilde{k}, n} = 2^{\frac{R}{\tau_1}} \left( \sigma^2 h_{se}^{\tilde{k}, n} + h_{sp}^{\tilde{k}} \left( \sigma^2 + \sum_{k \neq \tilde{k}} \bar{q}_s^k h_{se}^{k, n} + p_1 h_{pe}^n \right) \right)$$

$$- (\sigma^2 + p_1 h_{pp}) h_{se}^{\tilde{k}, n} - \left( \sigma^2 + \sum_{k \neq \tilde{k}} \bar{q}_s^k h_{se}^{k, n} \right) h_{sp}^{\tilde{k}}, \qquad (64)$$

$$C_{\tilde{k}, n} = 2^{\frac{R}{\tau_1}} \sigma^2 \left( \sigma^2 + \sum_{k \neq \tilde{k}} \bar{q}_s^k h_{se}^{k, n} + p_1 h_{pe}^n \right)$$

$$- (\sigma^2 + p_1 h_{pp}) \left( \sigma^2 + \sum_{k \neq \tilde{k}} \bar{q}_s^k h_{se}^{k, n} \right). \qquad (65)$$

*Proof*: Please refer to Appendix D. □

Once $p_s^{\tilde{k}}$ is obtained for each $\tilde{k} \in \mathbb{K}$, by searching over all $K+1$ possible SU schedulings, the solution to the problem in (41) is the one that maximizes the objective function in (51). The index of the SU scheduled for information transmission is obtained as $k^* = \arg\max_{\tilde{k} \in \mathbb{K}} f^{\tilde{k}}(p_s^{\tilde{k}}) - \eta \chi^{\tilde{k}}(p_s^{\tilde{k}})$, and we then have $p_s^{k^*}$ obtained from Theorem 2 and $q_s^{k^*} = 0, p_s^k = 0, q_s^k = \bar{q}_s^k, \forall k \neq k^*$. Note that if $k^* = 0$, then all the SUs are scheduled for jamming.

The proposed algorithm based on the dual optimization method and the BCD method is summarized in Algorithm 1.

*Remark 1*: The complexity of Algorithm 1 is analyzed as follows. Since the subgradient method used to update $\eta$ is

**Algorithm 1** Proposed joint SU scheduling, power and time allocation algorithm based on the dual optimization method and the BCD method.

1: Initialize $\eta(0)$, $t = 0$, $\mathbb{K} = \{0, \ldots, K\}$, and $\mathbb{N} = \{1, \ldots, N\}$.
2: **repeat**
3:    For each channel fading state, do the following steps 4-12.
4:    Initialize $\tau_1 = 0.5$, $p_1 = P_{max}$.
5:    **repeat**
6:       **for all** $\tilde{k}$ such that $\tilde{k} \in \mathbb{K}$ **do**
7:          For all $n \in \mathbb{N}$, calculate $p_s^{\tilde{k}, n}$ from (59).
8:          Calculate $n^*$ from (60) and obtain $p_s^{\tilde{k}} = p_s^{\tilde{k}, n^*}$.
9:       **end for**
10:      Calculate $k^*$ as $k^* = \arg\max_{\tilde{k} \in \mathbb{K}} f^{\tilde{k}}(p_s^{\tilde{k}}) - \eta \chi^{\tilde{k}}(p_s^{\tilde{k}})$, and let $q_s^{k^*} = 0, p_s^k = 0, q_s^k = \bar{q}_s^k$ for all $k \neq k^*$.
11:      Obtain $\tau_1$ from (33), and then obtain $p_1$ according to Theorem 1.
12:    **until** improvement of the objective function value in (30) converges to a prescribed accuracy.
13:    Update $\eta(t+1) = \eta(t) - \theta(t)(\varepsilon_0 - E\{\chi(\{p_s^k\}, \{q_s^k\}, p_1, \tau_1)\})$.
14:    $t = t+1$.
15: **until** $|\eta(t) - \eta(t-1)| \leq \epsilon$.
    where $\epsilon$ is the accuracy, $\theta(t)$ is a sequence of step sizes.

polynomial in the number of dual variables [47], its complexity is $\mathcal{O}(1)$. In each iteration of $\eta$, obtaining $\tau_1$, $\{p_s^k\}$, $\{q_s^k\}$ and $p_1$ requires $\mathcal{O}(KN\Xi)$ calculations, where $\Xi$ is the number of iterations for the BCD method to converge. Thus, the total complexity of Algorithm 1 is $\mathcal{O}(KN\Xi)$, which is linear in the number of SUs and the number of EAVs.

## V. EXTENSIONS

### A. Collusive EAVs

Here, we consider the EAVs are collusive and perfect CSI is available at the SUs. The case of unknown EAVs' CSI is discussed in Section V-C. It can be noted that the summation term in (9) makes the problem highly intractable. For this, considering the fact that at most one $p_s^k$ is nonzero, a lower bound on $C_{ws}$ in (10) denoted as $\tilde{C}_{ws}$ is proposed by ignoring the interference term $\sum_{k \in \mathbb{K}} p_s^k h_{se}^{k,n}$ in the second part of (9). Using $\tilde{C}_{ws}$, the PU secrecy outage probability $\tilde{\varepsilon}_{ps} = \Pr\left\{\tilde{C}_{ws} < R\right\}$ is an upper bound on $\varepsilon_{ps}$ and thus the PU secrecy constraint in (11) must be satisfied if $\tilde{\varepsilon}_{ps} \leq \varepsilon_0$ holds. By following the procedures used in Section IV, P1 with collusive EAVs can be solved with an algorithm similar to Algorithm 1. The details are omitted here due to page limit.

### B. Greedy Algorithms

In this subsection, we consider that the PU is greedy and requires the cooperative SUs to minimize the PU secrecy outage probability with perfect CSI available at the SUs. Based on this motivation, we formulate the optimization problem as

(P3)
$$\max_{\{p_s^k\},\{q_s^k\},p_0,p_1,\tau_0,\tau_1} E\left\{\tau_1 \sum_{k\in\mathbb{K}} \log_2\left(1 + \frac{p_s^k h_{ss}^k}{\sigma^2 + p_1 h_{psr}}\right)\right\}$$
$$- \eta\varepsilon_{ps} \quad (66)$$
$$\text{s.t. } (13)-(17), (19),$$

where the value of $\eta$ is a very large positive number. The second part of the objective function in (66) is introduced as a penalty function for punishing causing secrecy outage to the PU. It can be verified that Propositions 1-2 still hold for P3. Based on Propositions 1-2 and using the indicator function in (26), P3 can be reformulated and then decoupled into parallel subproblems with the same structure, each for one fading state as given by the problem in (30). Thus, using the BCD method adopted in Section IV, the solution for the case of perfect CSI with non-collusive EAVs can be obtained. As for the case of collusive EAVs, the method proposed in Sections V-A can be used to solve the problem. We omit the details here for brevity.

## C. Unknown EAVs' CSI

Here, this subsection investigates the unknown EAVs' CSI case. We assume that the instantaneous values as well as the distribution information of $h_{se}^{k,n}$ and $h_{pe}^n$ are unknown. In such a case, it is impossible to theoretically guarantee the PU secrecy constraint in (11). Instead, we design an efficient algorithm that can decrease the PU secrecy outage probability as much as possible in what follows. Firstly, in order to let the PU achieve a higher secrecy rate, the value of $p_0$ is set to zero and the value of $p_1$ is set to its maximum allowable value $\frac{P_{max}}{\tau_1}$. Secondly, in order to let the SU achieve a satisfactory performance, the SU with the maximum communication channel power gain to the CHAP is scheduled for information transmission in the WIT phase, i.e., $q_s^{k^*} = 0, p_s^k = 0, \forall k \neq k^*$, where $k^* = \arg\max_{k\in\mathbb{K}} h_{ss}^k$. Thirdly, in order to interfere with the EAVs to the most extent, the transmit power of the SU scheduled for jamming is set to the maximum allowable power, i.e., $q_s^k = \frac{\lambda h_{ss}^k Q\tau_0}{\tau_1}$ for all $k \neq k^*$. Then, the remaining variables $p_s^{k^*}, \tau_0, \tau_1$ are optimized to maximize the SU rate given by

$$\max_{p_s^{k^*}\geq 0, \tau_0\geq 0, \tau_1\geq 0} \tau_1 \log_2\left(1 + \frac{p_s^{k^*} h_{ss}^{k^*} \tau_1}{\sigma^2\tau_1 + P_{max}h_{psr}}\right) \quad (67)$$
$$\text{s.t. } \tau_0 + \tau_1 \leq 1, \quad (68)$$
$$p_s^{k^*}\tau_1 \leq \lambda h_{ss}^{k^*} Q\tau_0. \quad (69)$$

It can be verified that the optimal solution to the problem in (67) satisfies $\tau_0 + \tau_1 = 1$. It is proved briefly here. Suppose that the optimal solution satisfies $\tau_0 + \tau_1 < 1$. Then, the value of $\tau_0$ can be increased to satisfy $\tau_0 + \tau_1 = 1$ and such solution provides the same objective function value and satisfies the constraints. Thus, the optimal solution shall use up all the available time. It can be also verified that the optimal solution to the problem in (67) satisfies $p_s^{k^*}\tau_1 = \lambda h_{ss}^{k^*} Q\tau_0$, otherwise, we can increase $p_s^{k^*}$ to get a higher objective function value. Based on the above analysis, we have $\tau_0 = 1 - \tau_1$ and

$p_s^{k^*} = \frac{\lambda h_{ss}^{k^*} Q(1-\tau_1)}{\tau_1}$. Then, the problem in (67) reduces to the following problem as

$$\max_{0\leq \tau_1\leq 1} \tau_1 \log_2\left(1 + \frac{\lambda (h_{ss}^{k^*})^2 Q(1-\tau_1)}{\sigma^2\tau_1 + P_{max}h_{psr}}\right). \quad (70)$$

It can be verified that the above problem is non-linear and non-convex. Since $\tau_1$ is restricted in the narrow interval $[0,1]$, we can apply a one-dimensional search algorithm (such as the *fminbnd* function in MATLAB) within the interval $[0,1]$ to find the optimal $\tau_1$. The overall algorithm with unknown EAVs' CSI is summarized in Algorithm 2.

---
**Algorithm 2** Proposed algorithm with unknown EAVs' CSI.
1: Obtain $k^*$ as $k^* = \arg\max_{k\in\mathbb{K}} h_{ss}^k$.
2: Obtain $\tau_1$ by solving the problem in (70) via one-dimensional search algorithms and let $\tau_0 = 1 - \tau_1$.
3: Let $p_0 = 0$, $p_1 = \frac{P_{max}}{\tau_1}$, $p_s^{k^*} = \frac{\lambda h_{ss}^{k^*} Q(1-\tau_1)}{\tau_1}$, $q_s^{k^*} = 0$, and $p_s^k = 0$, $q_s^k = \frac{\lambda h_{ss}^k Q\tau_0}{\tau_1}$, for all $k \neq k^*$.

---

## VI. SIMULATION RESULTS

In this section, we verify the effectiveness of the proposed algorithms by Monte Carlo simulations. When performing simulations, the PTx is located at the center, the PRx is randomly deployed in a ring with inner radius 5 meter and outer radius 10 meter, while the CHAP, the SUs and the EAVs are randomly deployed in a ring with inner radius 10 meter and outer radius 20 meter. The channel power gains are modeled as $h = Lg$, where $L$ represents the large-scale pathloss modeled as $30 + 25\log_{10}(d)$ dB with $d$ denoting the distance, and $g$ is the normalized small-scale fading with exponential distribution. Besides, we set $\lambda = 0.5$, $R = 0.5$ bits/s/Hz, $Q = 10$ dBW, $P_{max} = 10$ dBW and $\sigma^2 = -90$ dBW. For the purpose of comparison, a benchmark algorithm without SU's cooperation is considered. The benchmark algorithm assumes that the PRx is wireless powered and supports full-duplex communication, and thus the PRx can send jamming signals to the EAV using the harvested energy by power splitting, similar to the one proposed in [30].

Fig. 2a and Fig. 2b show the SU ergodic rate and the PU secrecy outage probability against the required PU secrecy outage probability reduction $\triangle\varepsilon_p$, respectively. Since the greedy algorithm does not apply the PU secrecy constraint, the value of $\triangle\varepsilon_p$ has no impact on the SU ergodic rate and the PU secrecy outage probability achieved by the greedy algorithm. It is shown that, as $\triangle\varepsilon_p$ increases, the SU ergodic rate and the PU secrecy outage probability achieved by Algorithm 1 decrease and are higher than that achieved by the greedy algorithm. This is because that when dealing with the PU secrecy outage probability, Algorithm 1 aims to guarantee the reduction of the PU secrecy outage probability, while the greedy algorithm aims to minimize the PU secrecy outage probability. It is also shown that the greedy algorithm achieves significantly lower PU secrecy outage probability than the benchmark algorithm, and Algorithm 1 outperforms the benchmark algorithm if $\triangle\varepsilon_p$ is not too small and their performance gap increases as $\triangle\varepsilon_p$ increases. As expected, it is seen that the PU secrecy outage probability with collusive EAVs is higher than that with non-collusive EAVs. It is interesting to observe that the SU ergodic rate with collusive EAVs is higher than that with non-colllusive EAVs. This can be explained as follows. Since the PU secrecy outage probability with collusive EAVs is higher than that with non-colllusive EAVs, the SU will have more flexibility in cooperation with the PU for decreasing the PU secrecy outage probability. This indicates that it is beneficial to the SUs if the EAVs are collusive.

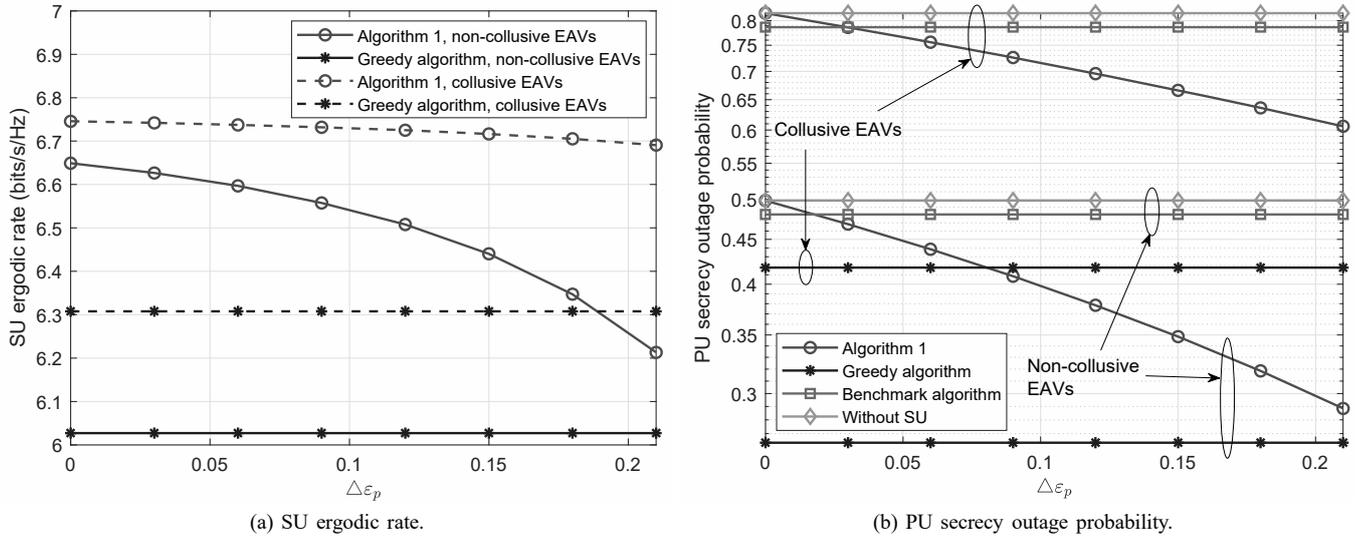

Figure 2. Performance against the secrecy outage probability reduction $\triangle\varepsilon_p$ ($K=100$, $N=8$).

Fig. 3a and Fig. 3b show the SU ergodic rate and the PU secrecy outage probability against the number of SUs $K$, respectively. It is shown that, with the increase of $K$, the SU ergodic rates achieved by Algorithm 1 and the greedy algorithm increase. This indicates that increasing the value of $K$ is beneficial to the SUs for both Algorithm 1 and the greedy algorithm. This is because that by increasing the number of SUs, there is a much better chance that the scheduled SU has more harvested energy and better channel conditions to achieve a higher rate. It is also shown that, with the increase of $K$, the PU secrecy outage probability achieved by Algorithm 1 keeps constant, while the PU secrecy outage probability achieved by the greedy algorithm decreases. This indicates that increasing the value of $K$ is beneficial to the PU only for the greedy algorithm. The above differences between the two algorithms lie in the fact that the aim of Algorithm 1 is to maximize the SU ergodic rate with guaranteed reduction of the PU secrecy outage probability while the aim of the greedy algorithm is to minimize the PU secrecy outage probability. Besides, it is shown that both Algorithm 1 and the greedy algorithm outperform the benchmark algorithm in terms of PU secrecy outage probability, and the gap between the greedy algorithm and the benchmark algorithm increases greatly as $K$ increases.

Fig. 4a and Fig. 4b show the SU ergodic rate and the PU secrecy outage probability against the number of EAVs $N$, respectively. It is shown that the PU secrecy outage probabilities achieved by Algorithm 1 and the greedy algorithm increase as $N$ increases. This is as expected since more EAVs leads to lower PU secrecy rate that can be achieved as observed from (8) and (10). It is also shown that with the increase of $N$, the SU ergodic rates achieved by both Algorithm 1 and the greedy algorithm increases. This indicates that a larger number of EAVs is actually beneficial to the SUs for both algorithms. This can be explained as follows. Since a larger $N$ leads to a higher PU secrecy outage probability, this can let the SUs cooperate with the PUs more flexibly and give the SUs more opportunities to achieve a higher SU ergodic rate.

Next, we investigate the performance of the proposed algorithm with unknown EAVs' CSI. Fig. 5a and Fig. 5b show the SU ergodic rate and the PU secrecy outage probability with unknown EAVs' CSI against the number of SUs $K$, respectively. By comparing Fig. 5a with Fig. 3a, it is seen that the SU ergodic rate with unknown EAVs' CSI drops significantly compared to the one with perfect CSI. From Fig. 5b, it is shown that the PU secrecy outage probability achieved by the proposed algorithm with unknown EAVs' CSI decreases as $K$ increases and is lower than that without SU when $K$ is not small. This indicates that as long as $K$ is sufficiently large, the PU security performance can be improved significantly even without knowlwdge of EAVs' CSI. Further in Fig. 6, we show the PU secrecy outage probability with unknown EAVs' CSI against the number of EAVs $N$. Note that, since the proposed algorithm with unknown EAVs' CSI does not depends on any information on EAVs, the achieved SU ergodic rate is constant as $N$ changes. It is shown that the PU secrecy outage probability achieved by the proposed algorithm with unknown EAVs' CSI increases as $N$ increases and is largerly lower than that without SU. From Fig. 5 and Fig. 6, we conclude that with unknown EAVs' CSI, the SU performance is greatly degraded while the PU security performance can still be improved largely by the proposed algorithm.

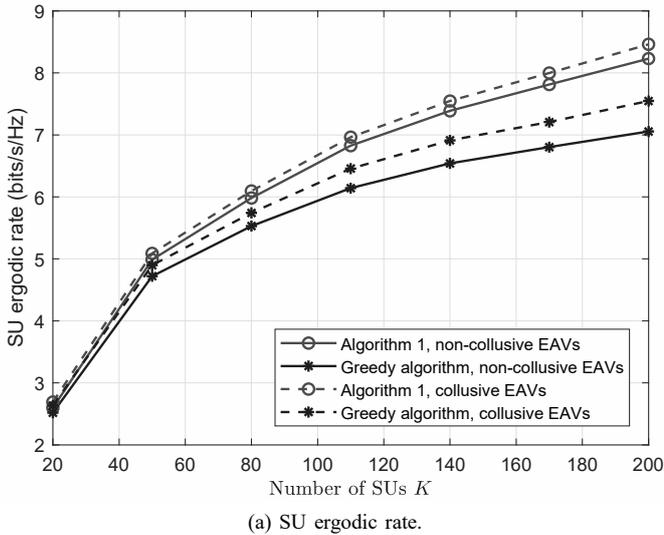
(a) SU ergodic rate.

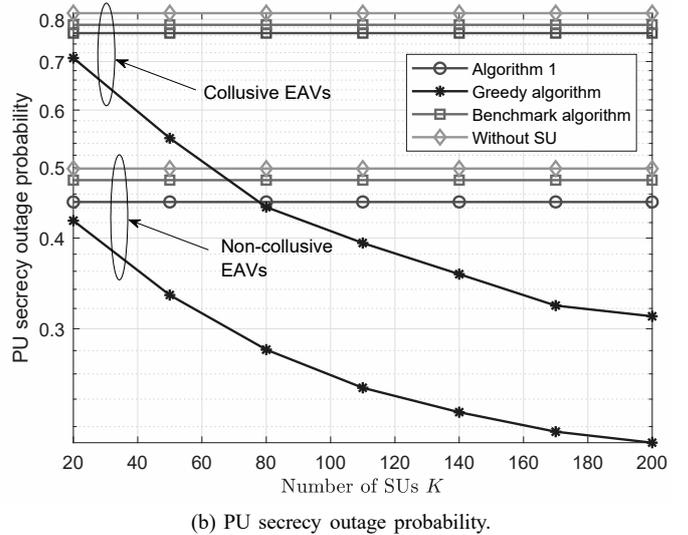
(b) PU secrecy outage probability.

Figure 3. Performance against the number of SUs $K$ ($\triangle\varepsilon_p = 0.1$, $N = 8$).

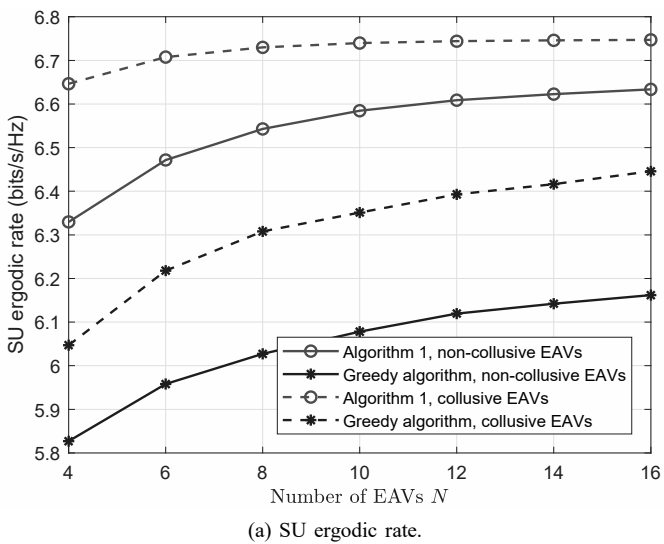
(a) SU ergodic rate.

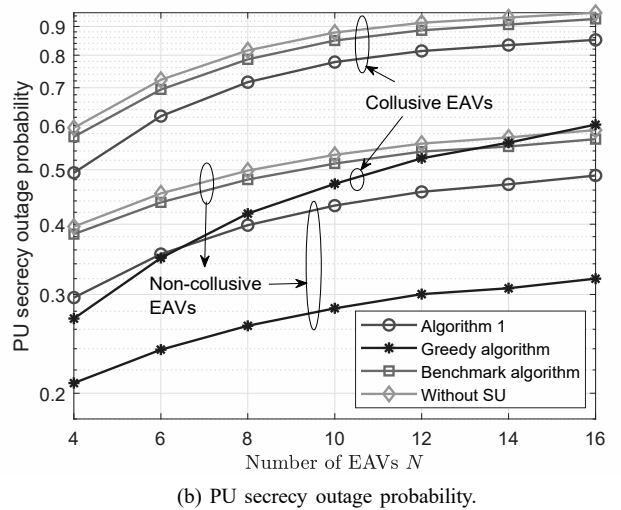
(b) PU secrecy outage probability.

Figure 4. Performance against the number of EAVs $N$ ($\triangle\varepsilon_p = 0.05$, $K = 100$).

## VII. CONCLUSIONS

This paper proposes a new cooperative protocol for the CWPCN to cooperate with the PU to improve the security performance of the PU who faces security threats from EAVs. In the protocol, the SUs first harvest energy and then use the harvested energy to interfere with the EAVs and gain their own transmission opportunities at the same time. Assuming that the CSI is perfect and the EAVs are collusive/non-collusive, we propose joint SU scheduling, power and time allocation algorithms based on the dual optimization method and the BCD method to maximize the SU ergodic rate under the PU secrecy constraint. In addition, we also propose more PU favorable greedy algorithms aimed at minimizing the PU secrecy outage probability. For the more practical unknown EAVs' CSI case, we furthermore propose an efficient algorithm for the SUs to improve the PU security performance.

Extensive simulations show that increasing the number of SUs is beneficial to both the SU and the PU, and increasing the number of EAVs degrades the PU security performance while actually improves the SU performance. It is also shown that, compared to non-collusive EAVs, collusive EAVs are more harmful to the PU while actually are more beneficial to the SUs. Besides, it is shown that with unknown EAVs' CSI, the SU performance is significantly degraded while the PU security performance can still be improved significantly by the proposed algorithm.

There are some possible extensions for the work in this paper. First, we have considered a single primary communication link in this paper, while it would be interesting to extend the results to support multiple primary links. In such a case, the SUs shall comply with the PU secrecy constraints for multiple primary communications links and the resource allocation shall achieve a balance among the requirements of different primary communication links. Second, we have considered that both the SU and the EAV are equipped with a single antenna, while equipping multiple antennas can enhance the performance of the SU and the EAV. In such a case, multiple antennas at the SU can let the SU interfere with the EAV more efficiently using beamforming technologies, while multiple antennas at the EAV can let the EAV eavesdrop the primary communication more conveniently using diversity combining technologies, and thus extending the results to this case is also interesting.

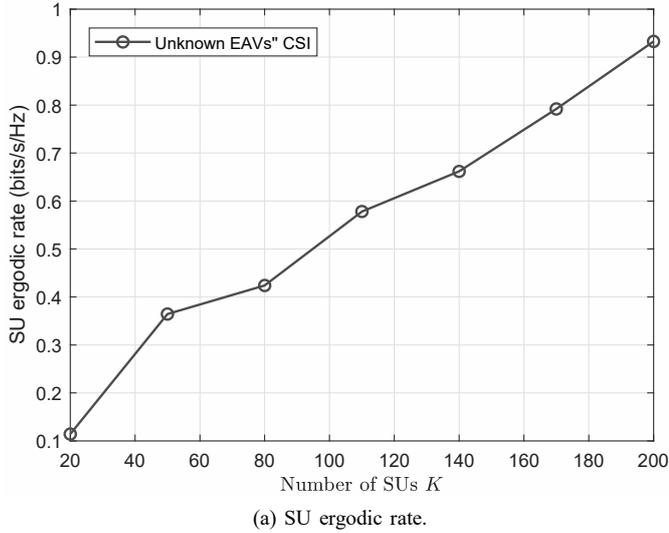

(a) SU ergodic rate.

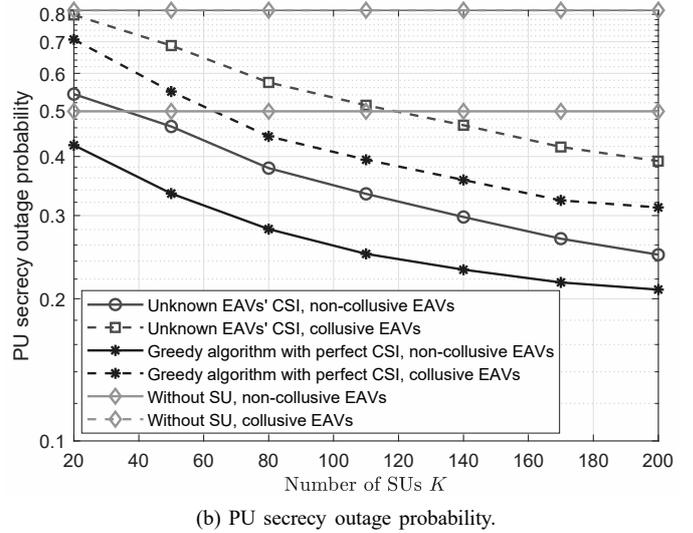

(b) PU secrecy outage probability.

Figure 5. Performance with unknown EAVs' CSI against the number of SUs $K$ ($N = 8$).

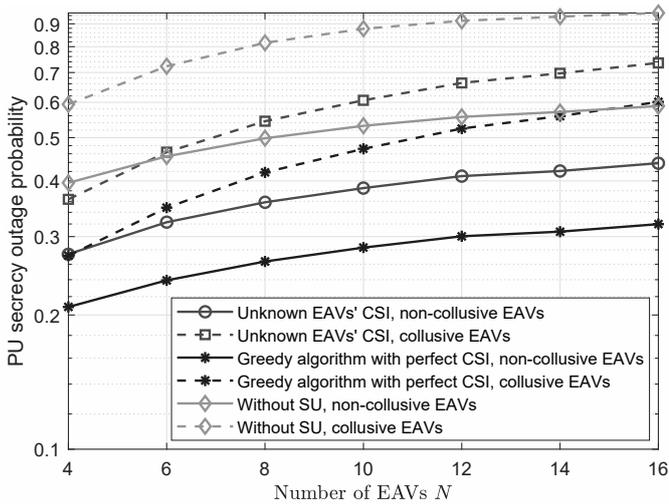

Figure 6. PU secrecy outage probability with unknown EAVs' CSI against the number of EAVs $N$ ($K = 100$).

## APPENDIX A
## PROOF OF PROPOSITION 1

Suppose that $\{\{\hat{p}_s^k\}, \{\hat{q}_s^k\}, \hat{p}_0, \hat{p}_1, \hat{\tau}_0, \hat{\tau}_1\}$ denote the optimal solution to P1 that satisfies $\hat{p}_0\hat{\tau}_0 + \hat{p}_1\hat{\tau}_1 < P_{max}$. Then, we consider another solution $\{\{\tilde{p}_s^k\}, \{\tilde{q}_s^k\}, \tilde{p}_0, \tilde{p}_1, \tilde{\tau}_0, \tilde{\tau}_1\}$ that satisfies $\{\tilde{p}_s^k\} = \{\hat{p}_s^k\}$, $\{\tilde{q}_s^k\} = \{\hat{q}_s^k\}$, $\tilde{p}_0 = \alpha_1 \hat{p}_0$, $\tilde{p}_1 = \hat{p}_1$, $\tilde{\tau}_0 = \hat{\tau}_0$, $\tilde{\tau}_1 = \hat{\tau}_1$, where $\alpha_1 > 1$. The value of $\alpha_1$ is chosen such that $\tilde{p}_0\tilde{\tau}_0 + \tilde{p}_1\tilde{\tau}_1 = P_{max}$. It can be easily seen that all the constraints of P1 are still satisfied by $\{\{\tilde{p}_s^k\}, \{\tilde{q}_s^k\}, \tilde{p}_0, \tilde{p}_1, \tilde{\tau}_0, \tilde{\tau}_1\}$. Thus, the solution $\{\{\tilde{p}_s^k\}, \{\tilde{q}_s^k\}, \tilde{p}_0, \tilde{p}_1, \tilde{\tau}_0, \tilde{\tau}_1\}$ is a feasible solution to P1. It can be seen that both $\{\{\hat{p}_s^k\}, \{\hat{q}_s^k\}, \hat{p}_0, \hat{p}_1, \hat{\tau}_0, \hat{\tau}_1\}$ and $\{\{\tilde{p}_s^k\}, \{\tilde{q}_s^k\}, \tilde{p}_0, \tilde{p}_1, \tilde{\tau}_0, \tilde{\tau}_1\}$ achieve the same objective function in (12). This means that $\{\{\tilde{\rho}_k\}, \{\tilde{p}_s^k\}, \tilde{p}_0, \tilde{p}_1, \tilde{\tau}_0, \tilde{\tau}_1\}$ is also an optimal solution to P1 and thus the optimal solution to P1 can always use up all the available transmit power of the PU. This completes the proof.

## APPENDIX B
## PROOF OF PROPOSITION 2

Suppose that $\{\{\hat{p}_s^k\}, \{\hat{q}_s^k\}, \hat{p}_0, \hat{p}_1, \hat{\tau}_0, \hat{\tau}_1\}$ is the optimal solution to P1 that satisfies $\hat{\tau}_0 + \hat{\tau}_1 < 1$. Then, we construct another solution $\{\{\tilde{p}_s^k\}, \{\tilde{q}_s^k\}, \tilde{p}_0, \tilde{p}_1, \tilde{\tau}_0, \tilde{\tau}_1\}$ that satisfies $\{\tilde{p}_s^k\} = \{\hat{p}_s^k\}$, $\{\tilde{q}_s^k\} = \{\hat{q}_s^k\}$, $\tilde{p}_0 = \frac{1}{\alpha}\hat{p}_0$, $\tilde{p}_1 = \hat{p}_1$, $\tilde{\tau}_0 = \alpha\hat{\tau}_0$, $\tilde{\tau}_1 = \hat{\tau}_1$, where $\alpha > 1$. The value of $\alpha$ is chosen such that $\tilde{\tau}_0 + \tilde{\tau}_1 = 1$. It can be easily verified that all the constraints of P1 still hold for $\{\{\tilde{p}_s^k\}, \{\tilde{q}_s^k\}, \tilde{p}_0, \tilde{p}_1, \tilde{\tau}_0, \tilde{\tau}_1\}$. Thus, $\{\{\tilde{p}_s^k\}, \{\tilde{q}_s^k\}, \tilde{p}_0, \tilde{p}_1, \tilde{\tau}_0, \tilde{\tau}_1\}$ is a feasible solution to P1. Since $\{\tilde{p}_s^k\} = \{\hat{p}_s^k\}$, $\tilde{p}_1 = \hat{p}_1$, $\tilde{\tau}_1 = \hat{\tau}_1$, the values of the objective function in (12) achieved by the two solutions are the same. This means that the optimal solution to P1 can always use up all the available unit time. This completes the proof.

## APPENDIX C
## PROOF OF THEOREM 1

After a few mathematical manipulations, the inequality $C_{ws}^n \geq R$ can be rewritten as

$$W_n p_1 \geq 2^{\frac{R}{\tau_1}} - 1, \qquad (71)$$

where $W_n = \frac{h_{pp}}{\sigma^2 + \sum_{k \in \mathbb{K}} p_s^k h_{sp}^k} - \frac{2^{\frac{R}{\tau_1}} h_{pe}^n}{\sigma^2 + \sum_{k \in \mathbb{K}} (p_s^k + q_s^k) h_{se}^{k,n}}$. Then, we can make the following discussions on the value of $\chi_n(\{p_s^k\}, \{q_s^k\}, p_1, \tau_1)$:

- *Case 1*: $W_n > 0$. In this case, the inequality $C_{ws}^n \geq R$ is reformulated as $p_1 \geq \frac{2^{\frac{R}{\tau_1}}-1}{W_n}$. Thus $\chi_n(\{p_s^k\},\{q_s^k\},p_1,\tau_1)$ equals to 0 if the inequality $p_1 \geq \frac{2^{\frac{R}{\tau_1}}-1}{W_n}$ holds and equals to 1 otherwise.
- *Case 2*: $W_n \leq 0$. In this case, considering the fact that $p_1 \geq 0$, the inequality $C_{ws}^n \geq R$ cannot be satisfied. Therefore, $\chi_n(\{p_s^k\},\{q_s^k\},p_1,\tau_1)$ equals to 1.

Let $p_1^n$ denote the optimal solution to the inner maximization problem in (38) for a given $n \in \mathbb{N}$. Based on the above discussions, we can make discussions on $p_1^n$ as:

- *Case 1*: $W_n > 0$. In this case, we discuss the value of $p_1^n$ in two subcases. In the subcase of $\frac{2^{\frac{R}{\tau_1}}-1}{W_n} > \bar{p}_1$, the value of $\chi_n(\{p_s^k\},\{q_s^k\},p_1,\tau_1)$ equals to 1 under the constraint (39), and thus $p_1^n = 0$. In the subcase of $\frac{2^{\frac{R}{\tau_1}}-1}{W_n} \leq \bar{p}_1$, the value $\frac{2^{\frac{R}{\tau_1}}-1}{W_n}$ is the critical value of $p_1$ over which the value of $\chi_n(\{p_s^k\},\{q_s^k\},p_1,\tau_1)$ changes from 1 to 0 under the constraint (39). Thus the inner maximization problem in (38) is maximized at two possible points as given by

$$p_1^n = \begin{cases} \frac{2^{\frac{R}{\tau_1}}-1}{W_n}, & \frac{2^{\frac{R}{\tau_1}}-1}{W_n} \leq \bar{p}_1, f_1\left(\frac{2^{\frac{R}{\tau_1}}-1}{W_n}\right) \geq f_1(0) - \eta, \\ 0, & \text{otherwise.} \end{cases} \quad (72)$$

- *Case 2*: $W_n \leq 0$. In this case, since $\chi_n(\{p_s^k\},\{q_s^k\},p_1,\tau_1)$ equals to 1 under the constraint (39), we can easily have $p_1^n = 0$.

The above discussions lead to the expression of $p_1^n$ given by

$$p_1^n = \begin{cases} \frac{2^{\frac{R}{\tau_1}}-1}{W_n}, & W_n > 0, \frac{2^{\frac{R}{\tau_1}}-1}{W_n} \leq \bar{p}_1, f_1\left(\frac{2^{\frac{R}{\tau_1}}-1}{W_n}\right) \\ & \geq f_1(0) - \eta, \\ 0, & \text{otherwise.} \end{cases} \quad (73)$$

Therefore, the optimal solution of the problem in (38) is given by $p_1 = p_1^{n^*}$, where $n^* = \arg\min_{n \in \mathbb{N}} f_1(p_1^n) - \eta \chi_n(\{p_s^k\},\{q_s^k\},p_1^n,\tau_1)$.

## APPENDIX D
## PROOF OF THEOREM 2

Thus, in what follows, we focus on solving the inner maximization problem in (57). For a given $n \in \mathbb{N}$, after a few mathematical manipulations, the inequality $C_{ws}^{\tilde{k},n} \geq R$ can be rewritten as $A_{\tilde{k},n}(p_s^{\tilde{k}})^2 + B_{\tilde{k},n}p_s^{\tilde{k}} + C_{\tilde{k},n} \leq 0$, where $A_{\tilde{k},n}$, $B_{\tilde{k},n}$ and $C_{\tilde{k},n}$ are given by (60), (64) and (65), respectively. Clearly, $A_{\tilde{k},n} > 0$. Then, we can make the following observations on the value of $\chi_n^{\tilde{k}}(p_s^{\tilde{k}})$:

- *Case 1*: $B_{\tilde{k},n}^2 - 4A_{\tilde{k},n}C_{\tilde{k},n} < 0$. In this case, the inequality $A_{\tilde{k},n}(p_s^{\tilde{k}})^2 + B_{\tilde{k},n}p_s^{\tilde{k}} + C_{\tilde{k},n} \leq 0$ cannot be satisfied and thus we have $\chi_n^{\tilde{k}}(p_s^{\tilde{k}}) = 1$.
- *Case 2*: $B_{\tilde{k},n}^2 - 4A_{\tilde{k},n}C_{\tilde{k},n} = 0$. In this case, the inequality $A_{\tilde{k},n}(p_s^{\tilde{k}})^2 + B_{\tilde{k},n}p_s^{\tilde{k}} + C_{\tilde{k},n} \leq 0$ is satisfied only if $p_s^{\tilde{k}} = -\frac{B_{\tilde{k},n}}{2A_{\tilde{k},n}}$. Thus we have $\chi_n^{\tilde{k}}(p_s^{\tilde{k}}) = 0$ if $p_s^{\tilde{k}} = -\frac{B_{\tilde{k},n}}{2A_{\tilde{k},n}}$ and $\chi_n^{\tilde{k}}(p_s^{\tilde{k}}) = 1$ otherwise.

- *Case 3*: $B_{\tilde{k},n}^2 - 4A_{\tilde{k},n}C_{\tilde{k},n} > 0$. In this case, the inequality $A_{\tilde{k},n}(p_s^{\tilde{k}})^2 + B_{\tilde{k},n}p_s^{\tilde{k}} + C_{\tilde{k},n} \leq 0$ is satisfied only if $x_{\tilde{k},n,1} \leq p_s^{\tilde{k}} \leq x_{\tilde{k},n,2}$, where $x_{\tilde{k},n,1} = \frac{-B_{\tilde{k},n} - \sqrt{B_{\tilde{k},n}^2 - 4A_{\tilde{k},n}C_{\tilde{k},n}}}{2A_{\tilde{k},n}}$ and $x_{\tilde{k},n,2} = \frac{-B_{\tilde{k},n} + \sqrt{B_{\tilde{k},n}^2 - 4A_{\tilde{k},n}C_{\tilde{k},n}}}{2A_{\tilde{k},n}}$. Thus, we have $\chi_n^{\tilde{k}}(p_s^{\tilde{k}}) = 0$ if $x_{\tilde{k},n,1} \leq p_s^{\tilde{k}} \leq x_{\tilde{k},n,2}$ and $\chi_n^{\tilde{k}}(p_s^{\tilde{k}}) = 1$ otherwise.

Based on the above discussions on $\chi_n^{\tilde{k}}(p_s^{\tilde{k}})$, for a given $n \in \mathbb{N}$, the optimal solution of the inner maximization problem in (57), $p_s^{\tilde{k},n}$, is discussed in the following three cases:

- *Case 1*: $B_{\tilde{k},n}^2 - 4A_{\tilde{k},n}C_{\tilde{k},n} < 0$. In this case, $\chi_n^{\tilde{k}}(p_s^{\tilde{k}}) = 1$ and thus the objective function $f^{\tilde{k}}(p_s^{\tilde{k}}) - \eta \chi_n^{\tilde{k}}(p_s^{\tilde{k}})$ becomes $f^{\tilde{k}}(p_s^{\tilde{k}}) - \eta$. Therefore, it is easy to check that $p_s^{\tilde{k},n} = \bar{p}_s^{\tilde{k}}$ due to the fact that $f^{\tilde{k}}(p_s^{\tilde{k}})$ is an increasing function of $p_s^{\tilde{k}}$.

- *Case 2*: $B_{\tilde{k},n}^2 - 4A_{\tilde{k},n}C_{\tilde{k},n} = 0$. In this case, we have $\chi_n^{\tilde{k}}(p_s^{\tilde{k}}) = 1$ only if $p_s^{\tilde{k}} \neq -\frac{B_{\tilde{k},n}}{2A_{\tilde{k},n}}$. Considering the constraint $0 \leq p_s^{\tilde{k}} \leq \bar{p}_s^{\tilde{k}}$, if $-\frac{B_{\tilde{k},n}}{2A_{\tilde{k},n}} < 0$ or $-\frac{B_{\tilde{k},n}}{2A_{\tilde{k},n}} > \bar{p}_s^{\tilde{k}}$, the objective function in (57) becomes $f^{\tilde{k}}(p_s^{\tilde{k}}) - \eta$ and thus $p_s^{\tilde{k},n} = \bar{p}_s^{\tilde{k}}$. Otherwise, under the condition $0 \leq -\frac{B_{\tilde{k},n}}{2A_{\tilde{k},n}} \leq \bar{p}_s^{\tilde{k}}$, the objective function in (57) is $f^{\tilde{k}}(p_s^{\tilde{k}}) - \eta$ if $p_s^{\tilde{k}} \neq -\frac{B_{\tilde{k},n}}{2A_{\tilde{k},n}}$ and is $f^{\tilde{k}}(p_s^{\tilde{k}})$ if $p_s^{\tilde{k}} = -\frac{B_{\tilde{k},n}}{2A_{\tilde{k},n}}$. Therefore, assuming that $0 \leq -\frac{B_{\tilde{k},n}}{2A_{\tilde{k},n}} \leq \bar{p}_s^{\tilde{k}}$, we have $p_s^{\tilde{k},n} = -\frac{B_{\tilde{k},n}}{2A_{\tilde{k},n}}$ if $f^{\tilde{k}}(-\frac{B_{\tilde{k},n}}{2A_{\tilde{k},n}}) \geq f^{\tilde{k}}(\bar{p}_s^{\tilde{k}}) - \eta$ and $p_s^{\tilde{k},n} = \bar{p}_s^{\tilde{k}}$ otherwise. In summary, we have

$$p_s^{\tilde{k},n} = \begin{cases} -\frac{B_{\tilde{k},n}}{2A_{\tilde{k},n}}, & 0 \leq -\frac{B_{\tilde{k},n}}{2A_{\tilde{k},n}} \leq \bar{p}_s^{\tilde{k}}, f^{\tilde{k}}(-\frac{B_{\tilde{k},n}}{2A_{\tilde{k},n}}) \\ & \geq f^{\tilde{k}}(\bar{p}_s^{\tilde{k}}) - \eta, \\ \bar{p}_s^{\tilde{k}}, & \text{otherwise.} \end{cases} \quad (74)$$

- *Case 3*: $B_{\tilde{k},n}^2 - 4A_{\tilde{k},n}C_{\tilde{k},n} > 0$. In this case, we have $\chi_n^{\tilde{k}}(p_s^{\tilde{k}}) = 0$ only if $x_{\tilde{k},n,1} \leq p_s^{\tilde{k}} \leq x_{\tilde{k},n,2}$. Considering the constraint $0 \leq p_s^{\tilde{k}} \leq \bar{p}_s^{\tilde{k}}$, the value of $p_s^{\tilde{k},n}$ is discussed in the following five subcases.
  - *Subcase 1*: $x_{\tilde{k},n,1} > \bar{p}_s^{\tilde{k}}$ or $x_{\tilde{k},n,2} < 0$. In this subcase, the objective function in (57) becomes $f^{\tilde{k}}(p_s^{\tilde{k}}) - \eta$ and thus we have $p_s^{\tilde{k},n} = \bar{p}_s^{\tilde{k}}$.
  - *Subcase 2*: $x_{\tilde{k},n,1} < 0$, $0 \leq x_{\tilde{k},n,2} \leq \bar{p}_s^{\tilde{k}}$. In this subcase, the objective function in (57) is $f^{\tilde{k}}(p_s^{\tilde{k}})$ if $0 \leq p_s^{\tilde{k}} \leq x_{\tilde{k},n,2}$ and is $f^{\tilde{k}}(p_s^{\tilde{k}}) - \eta$ if $x_{\tilde{k},n,2} < p_s^{\tilde{k}} \leq \bar{p}_s^{\tilde{k}}$. Therefore, since $f^{\tilde{k}}(p_s^{\tilde{k}})$ is an increasing function of $p_s^{\tilde{k}}$, the maximum value of the objective function in (57) is attained at $x_{\tilde{k},n,2}$ or $\bar{p}_s^{\tilde{k}}$, as given by

$$p_s^{\tilde{k},n} = \begin{cases} x_{\tilde{k},n,2}, & f^{\tilde{k}}(x_{\tilde{k},n,2}) \geq f_k(\bar{p}_s^{\tilde{k}}) - \eta, \\ \bar{p}_s^{\tilde{k}}, & \text{otherwise.} \end{cases} \quad (75)$$

  - *Subcase 3*: $x_{\tilde{k},n,1} < 0$, $x_{\tilde{k},n,2} > \bar{p}_s^{\tilde{k}}$. In this subcase, the objective function in (57) is $f^{\tilde{k}}(p_s^{\tilde{k}})$ and thus the maximum value of the objective function is attained at $p_s^{\tilde{k},n} = \bar{p}_s^{\tilde{k}}$.

- *Subcase 4*: $x_{\tilde{k},n,1} \geq 0$, $x_{\tilde{k},n,2} \leq \bar{p}_s^{\tilde{k}}$. In this subcase, the objective function in (57) is $f^{\tilde{k}}(p_s^{\tilde{k}})$ if $x_{\tilde{k},n,1} \leq p_s^{\tilde{k}} \leq x_{\tilde{k},n,2}$ and is $f^{\tilde{k}}(p_s^{\tilde{k}}) - \eta$ if $0 \leq p_s^{\tilde{k}} \leq x_{\tilde{k},n,1}$ or $x_{\tilde{k},n,2} < p_s^{\tilde{k}} \leq \bar{p}_s^{\tilde{k}}$. Therefore, since $f^{\tilde{k}}(p_s^{\tilde{k}})$ is an increasing function of $p_s^{\tilde{k}}$, the maximum value of the objective function in (57) is attained at $x_{\tilde{k},n,2}$ or $\bar{p}_s^{\tilde{k}}$, as given by

$$p_s^{\tilde{k},n} = \begin{cases} x_{\tilde{k},n,2}, & f^{\tilde{k}}(x_{\tilde{k},n,2}) \geq f^{\tilde{k}}(\bar{p}_s^{\tilde{k}}) - \eta, \\ \bar{p}_s^{\tilde{k}}, & \text{otherwise.} \end{cases} \quad (76)$$

- *Subcase 5*: $0 \leq x_{\tilde{k},n,1} \leq \bar{p}_s^{\tilde{k}}$, $x_{\tilde{k},n,2} > \bar{p}_s^{\tilde{k}}$. In this subcase, the objective function in (57) is $f^{\tilde{k}}(p_s^{\tilde{k}})$ if $x_{\tilde{k},n,1} \leq p_s^{\tilde{k}} \leq \bar{p}_s^{\tilde{k}}$ and is $f^{\tilde{k}}(p_s^{\tilde{k}}) - \eta$ if $0 \leq p_s^{\tilde{k}} < x_{\tilde{k},n,1}$. Thus, the maximum value of the objective function is attained at $p_s^{\tilde{k},n} = \bar{p}_s^{\tilde{k}}$.

Based on the above discussions, the optimal solution of the inner maximization problem in (57), $p_s^{\tilde{k},n}$, is summarized by expression (59). Then, the optimal solution of the problem in (57) is given by $p_s^{\tilde{k}} = p_s^{\tilde{k},n^*}$, where $n^* = \arg\min_{n \in \mathbb{N}} f^{\tilde{k}}(p_s^{\tilde{k},n}) - \eta\chi_n^{\tilde{k}}(p_s^{\tilde{k},n})$. This completes the proof.